\documentclass{emulateapj}
\usepackage{comment}
\usepackage{natbib}
\usepackage{xcolor}
\usepackage{amsmath}
\usepackage[backref,breaklinks,colorlinks,citecolor=blue]{hyperref}

\begin{document}

\title{Matching the dark matter profiles of dSph galaxies with those of simulated satellites: a two parameter comparison}

\author{Maarten A. Breddels\altaffilmark{1}, Carlos Vera-Ciro\altaffilmark{2} and
  Amina Helmi\altaffilmark{1} }

\altaffiltext{1}{Kapteyn Astronomical Institute, University of Groningen, P.O. Box 800, 9700 AV Groningen, The Netherlands}

\altaffiltext{2}{Department of Astronomy, University of Wisconsin, 2535
  Sterling Hall, 475 N. Charter Street, Madison, WI 53076, USA}

\email{e-mail:breddels@astro.rug.nl}

\newcommand{\mint}{\ensuremath{{\rm{min}}}}
\newcommand{\model}[1]{\texttt{#1}}
\newcommand{\light}{\ensuremath{{\rm{light}}}}
\newcommand{\data}{\ensuremath{{\rm{data}}}}
\newcommand{\kin}{\ensuremath{{\rm{kin}}}}
\newcommand{\nudm}{\ensuremath{\nu_\dm}}
\newcommand{\Msol}{\ensuremath{\text{M}_{\odot}}}
\newcommand{\Lsol}{{\rm L}_{\odot}}
\newcommand{\kms}{\ensuremath{{\rm km\,s}^{-1}}\xspace}

\newcommand{\Gammathree}{\ensuremath{\gamma_{-3}}}
\newcommand{\Mthree}{\ensuremath{M_{-3}}}

\newcommand{\edit}[1]{\textcolor{orange}{#1}}

\begin{abstract}
  We compare the dark matter halos' structural parameters derived for
  four Milky Way dwarf spheroidal galaxies to those of subhalos found
  in cosmological $N$-body simulations. We confirm that estimates of
  the mass at a single fixed radius are fully consistent with the
  observations.  However, when a second structural parameter such as
  the logarithmic slope of the dark halo density profile measured
  close to the half-light radius is included in the comparison, we
  find little to no overlap between the satellites and the
  subhalos. Typically the right mass subhalos have steeper profiles at
  these radii than measurements of the dSph suggest.  Using energy
  arguments we explore if it is possible to solve this discrepancy by
  invoking baryonic effects.  Assuming that feedback from supernovae
  can lead to a reshaping of the halos, we compute the required
  efficiency and find entirely plausible values for a significant
  fraction of the subhalos and even as low as 0.1\%.  This implies
  that care must be taken not to exaggerate the effect of supernovae
  feedback as this could make the halos too shallow. These results
  could be used to calibrate and possibly constrain feedback recipes
  in hydrodynamical simulations.

\end{abstract}

\keywords{galaxies: dwarf -- galaxies: kinematics and dynamics}

\section{Introduction}

The $\Lambda$CDM currently favored paradigm of structure formation has
proven to be a successful model to explain the evolution and large
scale structure of the Universe. It is on small scales, however, where
the concordance cosmology faces most of its challenges. One example is
the well-known mismatch in the number of observed satellites around
systems like the Milky Way and those found in cosmological $N$-body
simulations \citep{Klypin1999, Moore1999}.  Although this can be
solved by invoking baryonic effects, a new problem has been reported
regarding the abundance of massive satellites \citep[the
so-called ``Too Big to Fail problem'',][]{BoylanKolchin2011}.  Pure
$N$-body simulations in $\Lambda$CDM also predict self-similar halos
that have divergent central densities $\rho \sim r^{-1}$ \citep{NFW1,
  NFW2}. However, models of the internal kinematics of Milky Way dSph
have so far yielded inconclusive results regarding the inner slopes of
the halos hosting these galaxies \citep{Battaglia2008, Walker2009,
  Breddels2013four, Amorisco2014}.

The fundamental question that these studies have tried to address is
whether a cusped dark matter density profile fits the observations
better than a cored one. This is however, very difficult to answer
with current data. For instance, \citet{Breddels2013} using
non-parametric orbit-based dynamical models showed that, although
small values of the logarithmic inner slope of the dark matter density
profile are preferred in the case of the Sculptor dSph, the inner
slope value is strongly degenerate with the scale radius and the
uncertainties are large. A more feasible task is to measure the
logarithmic slope at a finite radius, e.g. at $r_{-3}$, the radius
where the logarithmic slope of the {\it light} profile is
$-3$. \citet[][hereafter BH13]{Breddels2013four} have determined the
logarithmic slope robustly (i.e. independently of the assumed profile,
be it cored or cusped) at $r_{-3}$, for the dSph Fornax (Fnx) and
Sculptor (Scl), and placed strong constraints on the values for
Sextans (Sxt) and Carina (Car). In addition, BH13 confirmed using
Schwarzschild modeling, that the mass at this radius can also be
tightly constrained \citep{Strigari2008, Walker2009, Wolf2010MNRAS}.

This result shifts the discussion to a better posed question, namely:
Are the logarithmic slope and mass at $r_{-3}$ inferred for these four
dwarfs consistent with those found in simulations? This phrasing has
the additional advantage that measuring these two quantities in
simulations of dwarf galaxies is less prone to numerical
artifacts. In this Letter we focus on answering this question using
the Aquarius simulations \citep{Springel2008}. Since these simulations
do not include baryons, discrepancies could possibly be attributed to
missing physics. For instance, early adiabatic contraction could make
the density distribution of the central parts steeper
\citep{Blumenthal1986}. On the other hand, feedback by supernovae
(SNe) may modify the dark matter profile and lead to a shallower inner
slope, especially for low mass galaxies \citep{Navarro1996,
  Mashchenko2006, Mashchenko2008, Governato2012, Ponzen2012MNRAS,
  Teyssier2013, diCintio2014b, diCintio2014}. Finally, dynamical
friction of large baryonic clumps has also been shown to modify the
logarithmic slope in the same sense \citep{El-Zant2001,
  Goerdt2006, Nipoti2015}.

This Letter is structured as follows. In Section \ref{sec:comparison}
we compare the observational results from BH13 to
the subhalos from the Aquarius simulations and highlight the
discrepancies. In Section \ref{sec:transform} we use analytic
arguments to estimate the minimum amount of energy needed to transform
the subhalos' profiles to a distribution more
consistent with the observational constraints following the arguments
laid out by \cite{Penarrubia2012,Amorisco2014,Maxwell2015}, and
derive the required SNe feedback efficiencies. Finally in Section
\ref{sec:conclusions} we discuss the broader implications of our results.
\\

\section{Comparison of the dark halos structural properties}
\label{sec:comparison}
\newcommand{\scalefnx}{0.75}
\begin{figure*}[t]
  \begin{center}
    \begin{tabular}{c c}
      \includegraphics[scale=\scalefnx]{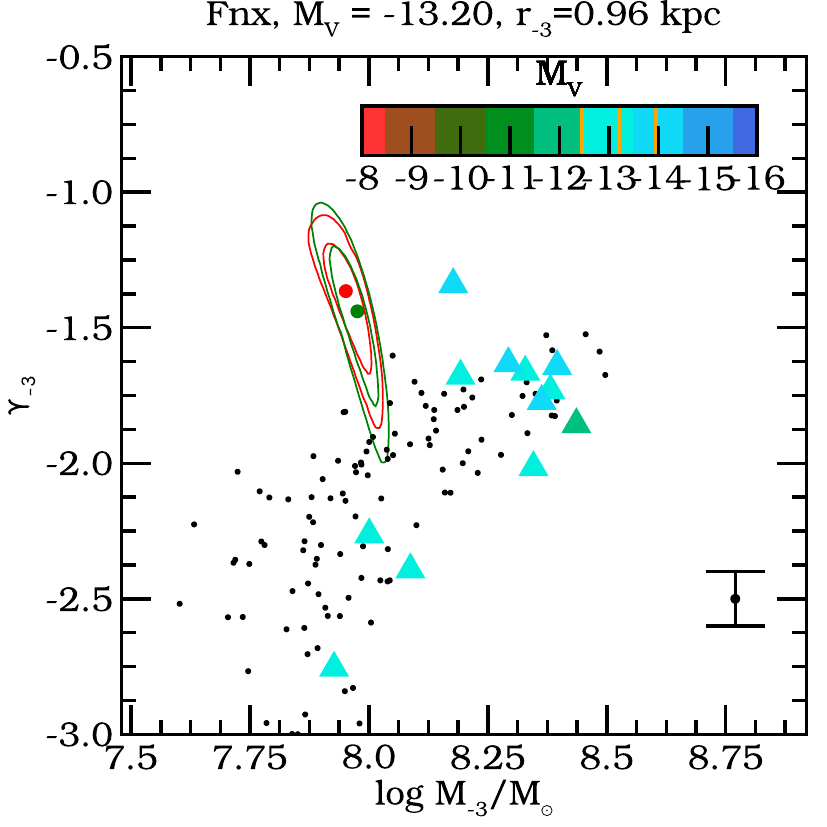} & \includegraphics[scale=\scalefnx]{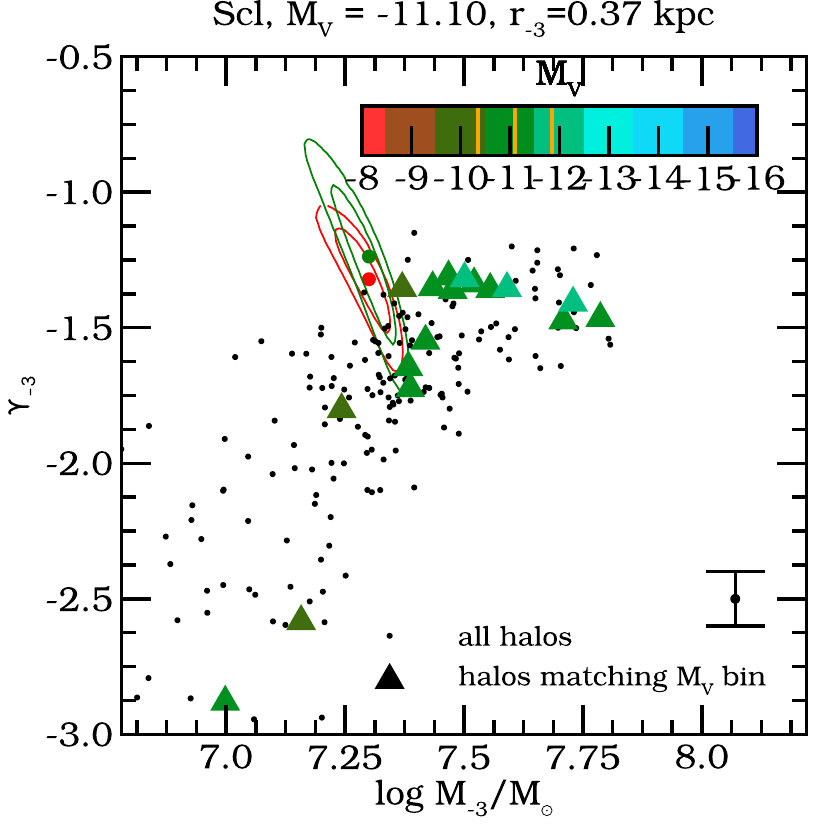} \\
      \includegraphics[scale=\scalefnx]{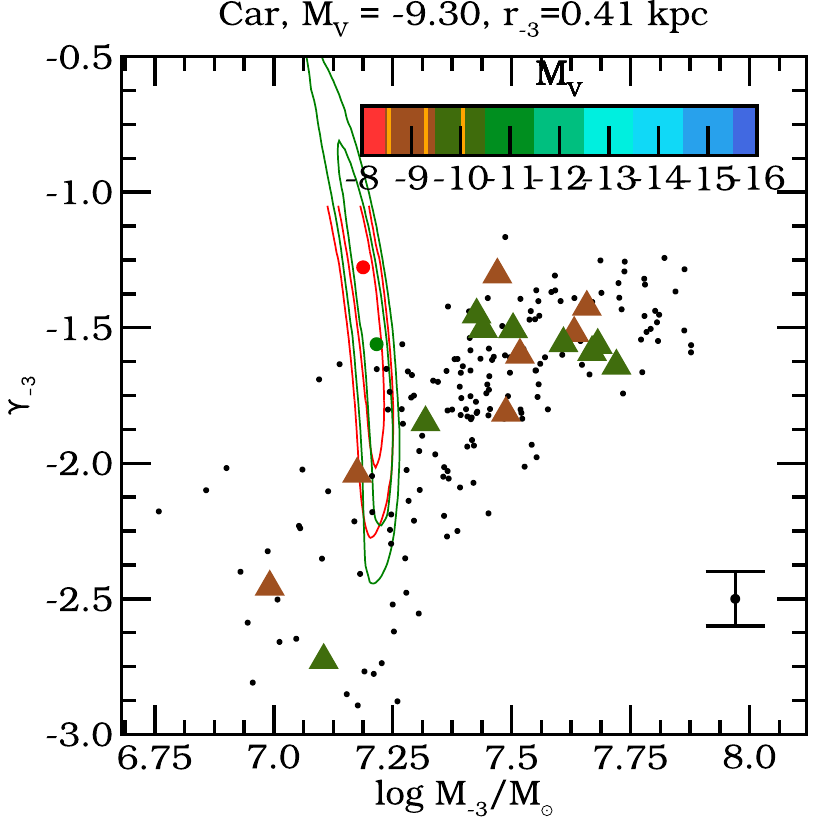} & \includegraphics[scale=\scalefnx]{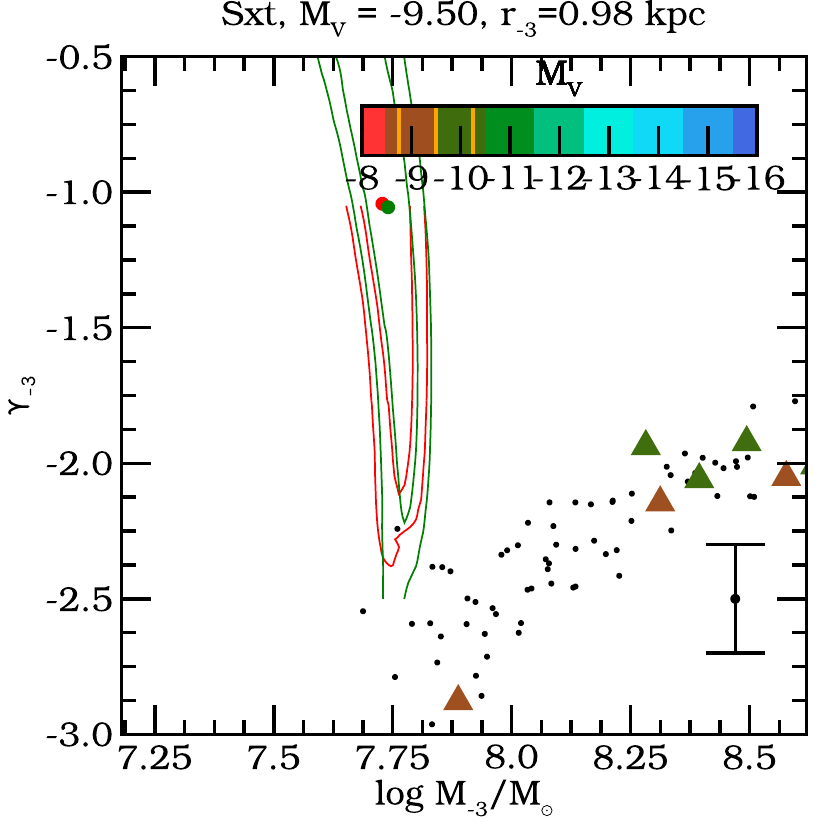}
    \end{tabular}
    \caption{Joint probability distribution function for the
      logarithmic slope \Gammathree{} and mass \Mthree{} of the dark
      matter halo at $r_{-3}$ for our sample of dSph. The contours
      represent the $1\sigma$ and $2\sigma$-equivalent confidence
      regions for an \model{NFW} profile (red, Eq.~\eqref{eq:nfw})
      and the \model{core13} model (green, Eq.~\eqref{eq:core}). The black
      points correspond to the subhalos of
      the six scaled Aquarius Milky Way-like halos, while the colored
      triangles are the subset with luminosity within $\pm
      0.75$ of the value of the respective dSph as predicted by the
      semi-analytic model of \citet{Starkenburg2013}. The error bar on the
      bottom right indicates the uncertainty in the slope estimated
      for the simulations.}
    \end{center}
\label{fig:m3slopes}.
\end{figure*}

The Aquarius simulations are a suite of six (labeled Aq-A to Aq-F)
high resolution cosmological dark matter only runs of Milky Way-like
halos. Here we use the highest resolution simulations available for
all six halos (known as level-2), with a particle mass of $\sim1.4
\times 10^4\Msol$ and a softening length of $66$ pc \citep[for more
details see][]{Springel2008}. For a more direct comparison of the
different systems, we have scaled all halos such that the Milky
Way-like hosts have a common mass of $M_{\rm halo} =
8.2\times10^{11}\Msol$ \citep[as in][]{VeraCiro2013}.  A dark halo
potentially hosting a galaxy such as Fnx ($M_{\rm vir} \sim 10^{9.5}
\Msol$) is resolved with $\sim 2 \times 10^5$ particles, and we expect
our results to be reliable beyond a radius of $\sim 200$~pc, which for
Fnx corresponds to $\sim 0.2 \,r_{-3}$.

As mentioned in the Introduction, BH13 have shown that, besides
\Mthree{} \citep[the mass within $r_{-3}$,][]{Strigari2008,
  Walker2009, Wolf2010MNRAS}, also $\Gammathree= \left.{\rm d} \log
  \rho_{\rm dm} /{\rm d} \log r \right| _{r_{-3}}$ can be robustly
determined for the dSph using current data \citep[see
also][]{WolfBullock2012}. \Gammathree{} does not depend strongly on
the assumed parametric profile that is used, whether cusped, cored or
logarithmically cored\footnote{A distinction is made between strictly
  cored ($\lim_{r\to 0} {\rm d} \rho_{\rm dm} / {\rm d}r$ $= 0$), and
  logarithmically cored ($\lim_{r\to 0} {\rm d} \log \rho_{\rm dm} /
  {\rm d} \log r = 0 $).}. These same two quantities can also be
estimated for our simulated halos. For each Aquarius satellite we
compute both \Mthree{} and \Gammathree{} using the values of $r_{-3}$
measured for each individual dSph galaxy. \Mthree{} follows trivially
from the number of particles found within $r_{-3}$. We determine the
slope $\gamma$ numerically using the algorithm described in
\citep{Lindner2015}. The values obtained are in good agreement with
those derived assuming parametric forms for the density profile (such
as truncated Einasto or \model{NFW} profiles), which are fitted to the
data and from which $\gamma$ can then be computed. Comparing the
non-parametric and parametric estimates we gauge the uncertainty on
$\gamma$ to be $\sim 0.2$.

In Fig. \ref{fig:m3slopes}, we show the posteriors for the joint
distribution of \Mthree{} and \Gammathree{} derived by
BH13 separately for each dSph galaxy. The red contours
correspond to the 1 and 2 $\sigma$ equivalent contours for the \model{NFW}
\citep{NFW1, NFW2} profile
\begin{equation}\label{eq:nfw}
\rho_{\rm dm}(r) = \rho_{0} \frac{1}{r/r_s(1+ r/r_s)^2},
\end{equation}
while green corresponds to the logarithmic cored profile
(\model{core13} in the nomenclature of BH13)
\begin{equation}\label{eq:core}
\rho_{\rm dm}'(r) = \rho_{0}' \frac{1}{(1+ r/r_s')^3},
\end{equation}
where $r_s$ and $r_s'$ are the scale radii and $\rho_0$ and $\rho_0'$
characteristic densities. We also plot in these panels the
subhalos' structural parameters determined non-parametrically and
computed at the $r_{-3}$ of the corresponding dSph. We show here only
those subhalos that host a luminous galaxy according to the
semi-analytic model of \citet{Starkenburg2013}.  The black dots
correspond to all such subhalos present in the six scaled Aquarius
simulations, while the colored triangles are those that host a galaxy
with an absolute magnitude within $\pm 0.75$ of the respective dSph.

From the top left panel, we see that the subhalos overlap little with
Fnx, especially if only the correct luminosity range is
considered. Note that if one only compares the \Mthree{} values, then
some subhalos with the right luminosity do actually
match. Also if only the slope \Gammathree{} were
to be taken into account, several subhalos would be compatible with
the observational constraints.

However the discrepancy becomes noteworthy for Fnx when comparing the
observations and simulations in the two-dimensional space spanned by
\Mthree{} and \Gammathree{}. For Scl and Car there is some, although
small, amount of overlap with the subhalos, but this is
less so for Sxt. Both Car and Sxt have large uncertainties on the
slope estimate, which makes the comparison less conclusive.

Therefore it is clear that these four dwarfs all show a
systematic deviation from the simulations: the slope of their dark
matter profiles are shallower than those found in the Aquarius
simulations, even if subhalos with a broader range of luminosities
are considered.

\section{Reshaping dark matter halos}
\label{sec:transform}

\begin{figure}[t]
  \begin{center}
    \begin{tabular}{c}
      \includegraphics[scale=\scalefnx]{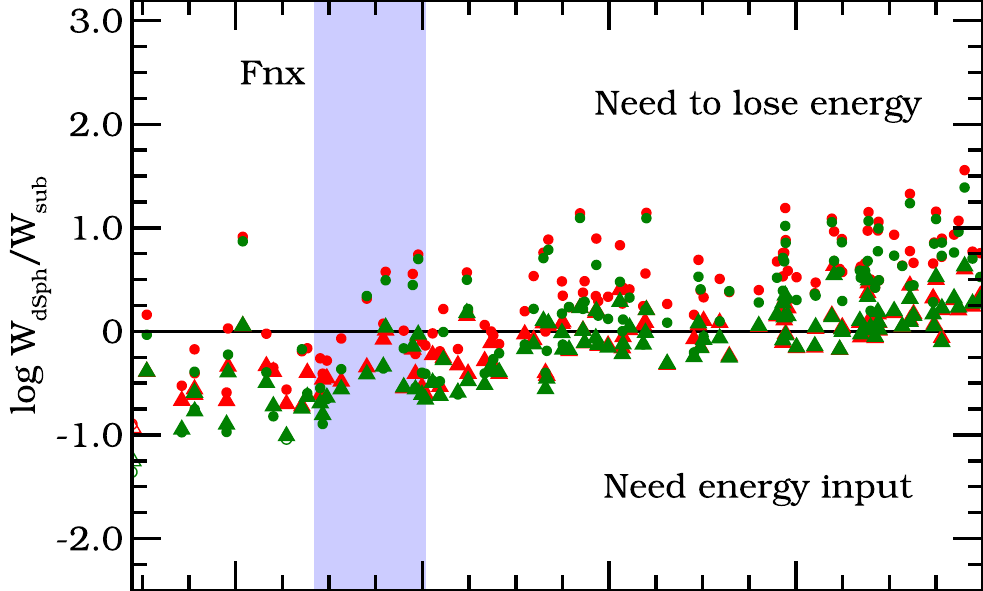} \\
      \includegraphics[scale=\scalefnx]{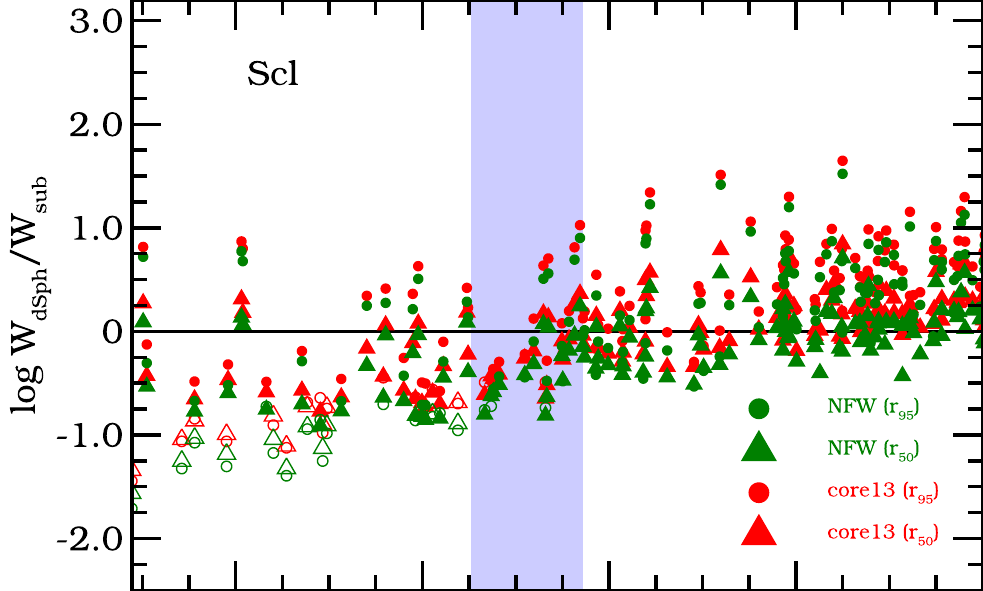} \\
      \includegraphics[scale=\scalefnx]{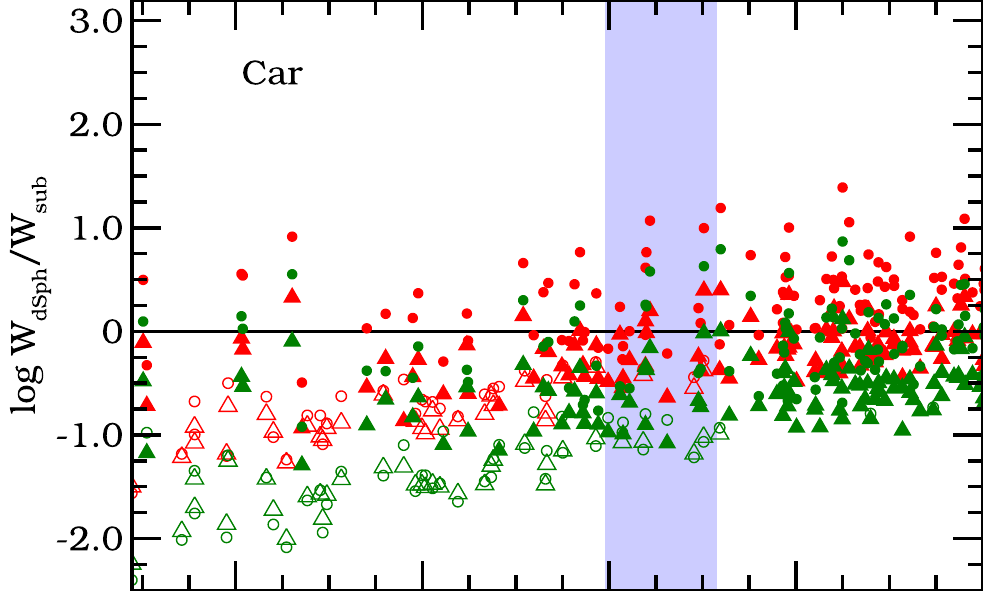} \\
      \includegraphics[scale=\scalefnx]{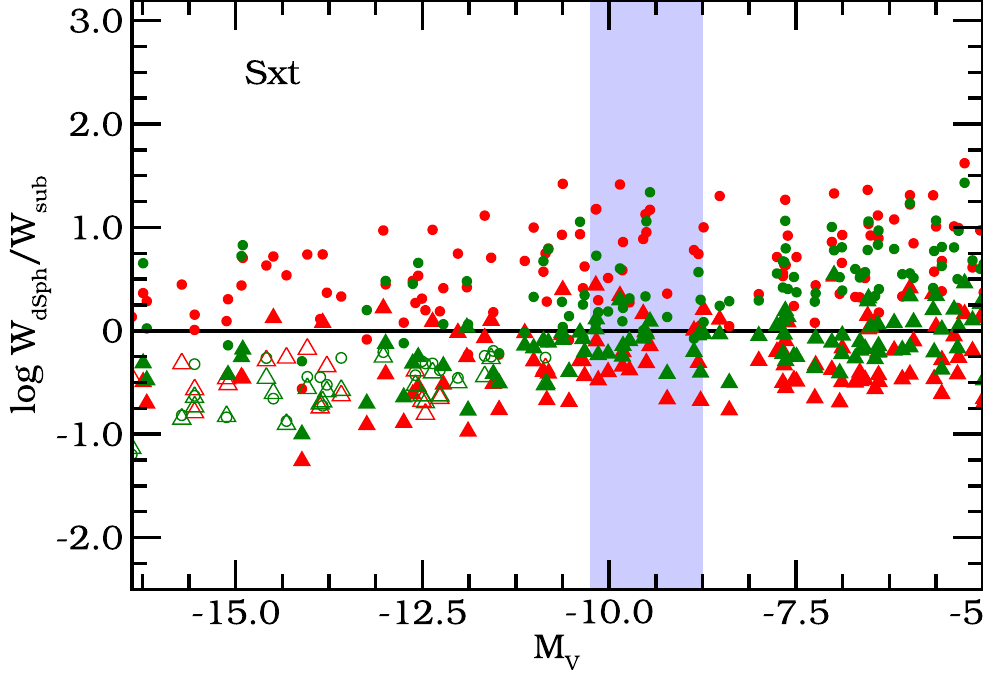}
    \end{tabular}
    \caption{Ratio of the potential energy of the dSph, $W_{\rm
        dSph}$, to that of the Aquarius subhalos $W_{\rm sub}$, as a
      function of their predicted absolute magnitude $M_V$, computed
      for $r_{95}$ (circles) and $r_{50}$ (triangles) for the
      \model{NFW} (red) and \model{core13} (green) models. The blue
      band indicates an absolute magnitude range within $\pm 0.75$ of
      the of the corresponding dSph. The open symbols are subhalos
      which cannot be transformed with the energy available in
      SNe. \label{fig:energies}}
  \end{center}
\end{figure}

\begin{figure}[t]
  \begin{center}
    \begin{tabular}{c}
      \includegraphics[scale=\scalefnx]{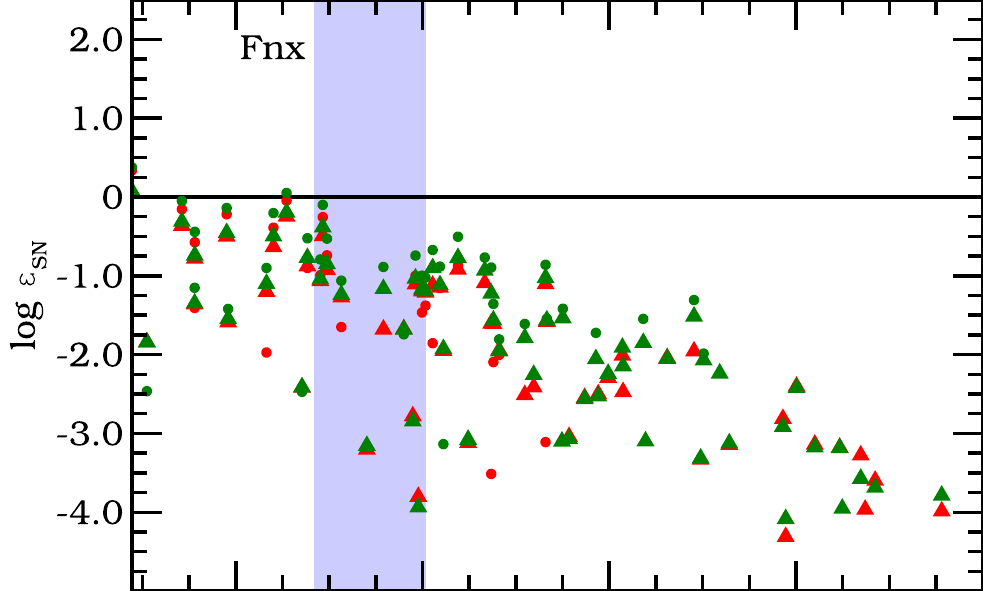} \\
      \includegraphics[scale=\scalefnx]{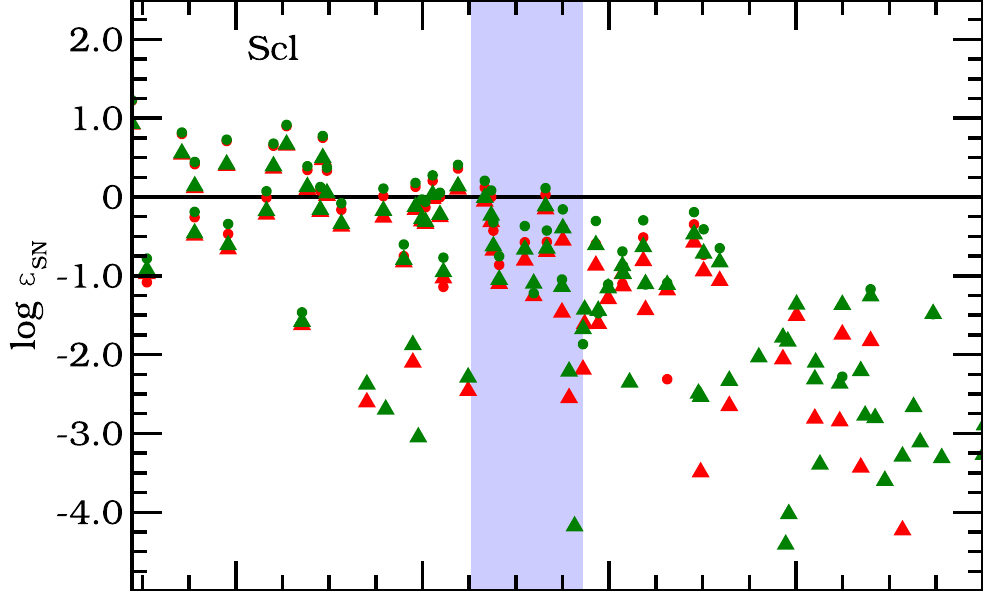} \\
      \includegraphics[scale=\scalefnx]{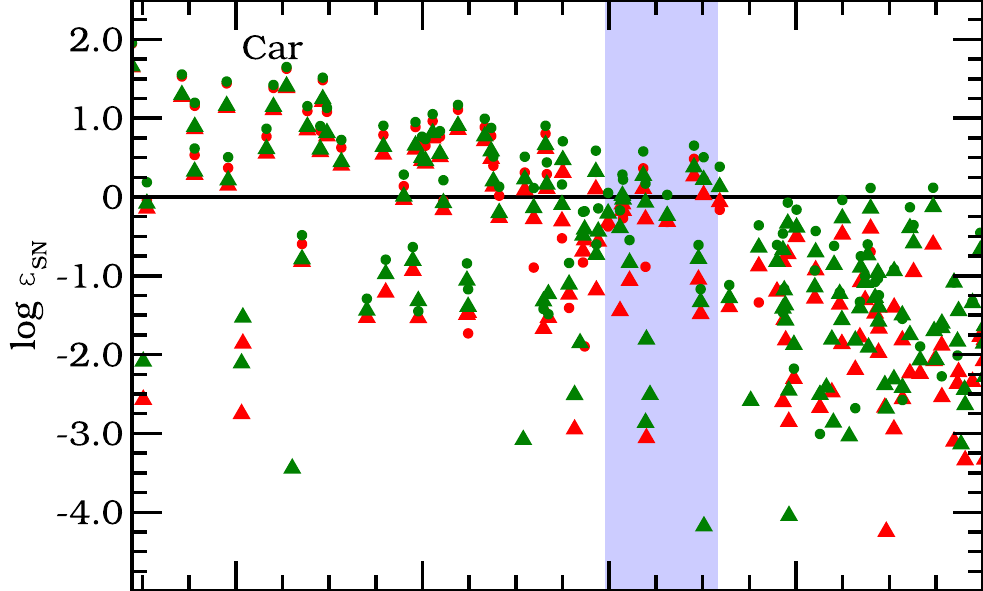} \\
      \includegraphics[scale=\scalefnx]{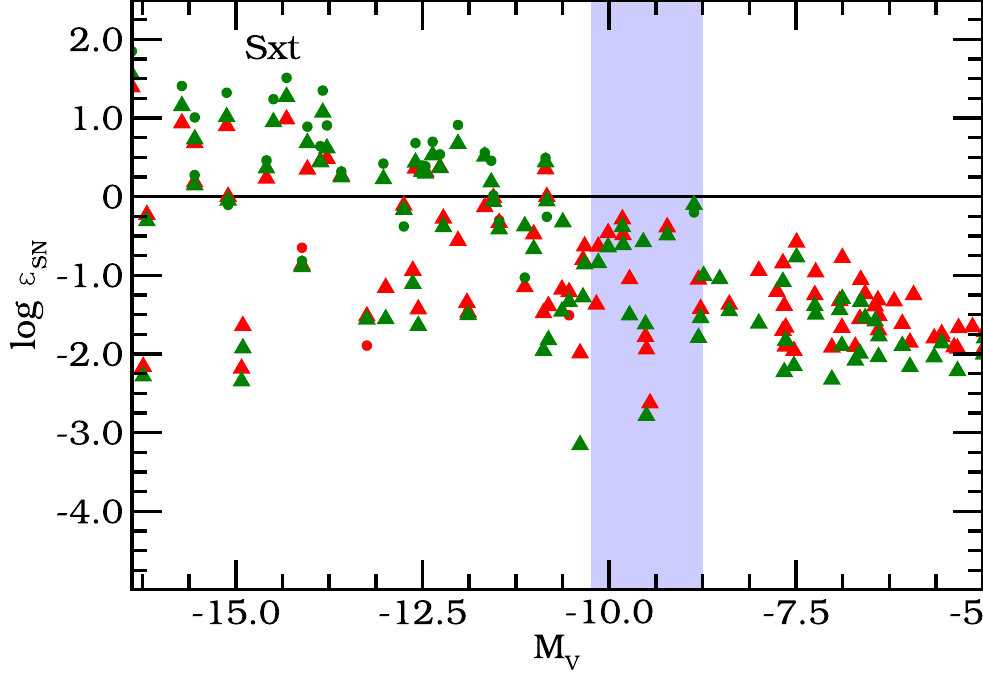}
    \end{tabular}
    \caption{Similar to Fig \ref{fig:energies}, now showing the SN
      efficiency needed to transform the Aquarius halos. \label{fig:SN_eff}}
  \end{center}
\end{figure}

As we have just seen dark matter only simulations in the $\Lambda$CDM
cosmology, predict steeper slopes of the dark matter profiles than
those determined for the dSphs.  The most commonly proposed mechanism
to transform the shape of the dark matter profile is feedback from SNe
that affects the baryons and, through gravity, ultimately alters the
dark matter mass distribution in these systems \citep{Mashchenko2006,
  Mashchenko2008, Governato2012, Ponzen2012MNRAS, Teyssier2013,
  diCintio2014}.

To establish whether a given Aquarius subhalo can be reshaped into a
halo that is compatible with a dSph, we first need to know how much
energy is required to do the transformation. Following
\citet{Penarrubia2012} we assume that each dwarf is in virial
equilibrium, and we compute the energy needed to transform a subhalo
$i$ to match the dSph model $j$ (given by Eq.~\eqref{eq:nfw} and
Eq.~\eqref{eq:core}): $\delta{}E = \left(W_{\rm dSph} - W_{\rm sub}
\right)/2$. Here $W_{\rm sub}$ is the potential energy of the subhalo
and $W_{\rm dSph}$ is the potential energy of the best fitting model
for our dSph:
\begin{equation}
  W_{\rm dSph} = -4 \pi G \int_0^{r_{\max,i}} {\rm d}r \rho_{\text{dm},j}(r) M_{j} (r)r,
\end{equation}
with $M(r)= M_{\rm dm}(r)$, i.e. we ignore the baryonic component
which is sub-dominant for dSph.  We choose two different values of
$r_{\max}$, namely $r_{50}$ and $r_{95}$, the radii enclosing 50\% or
95\% of the bound mass of each subhalo. The potential energy at
$r_{95}$ is more negative than at $r_{50}$, and thus the energy
required to transform the system inside $r_{50}$ is smaller than that
required for $r_{95}$. (Note that these values are larger than those
used by \citet{Maxwell2015}, who took $r_{\max} = 3 r_c $, with $r_c$
the radius where the dark halo profile's logarithmic slope is $-0.5$).

We calculate $W_{\rm sub}$ directly using the particles from the dark
matter simulations:
\begin{eqnarray}
 W_{\rm sub} &=& -4 \pi G \int_0^{r_{\max,i}}  {\rm d}r\rho_{\text{\rm dm},i}(r) M_{i} (r)r \nonumber \\
 &=& -G \int_0^{r_{\max, i}}  {\rm d}M M_{i}(r)/r \nonumber \\
 &=& -G \sum_{k=1}^N m^2 \,\, k/r_k,
\end{eqnarray}
where $m$ is the dark matter particle mass, $r_k$ is the sorted
radial distance of particle $k$ from the center of the subhalo such
that $r_1$ is the innermost particle and $r_N$ that
farthest away but still within $r_{\max}$.

Fig.~\ref{fig:energies} shows for each dSph the ratio $W_{\rm
  dSph}/W_{\rm sub}$ as a function of absolute magnitude $M_V$ as
given by the semi-analytic model. The blue bands indicate a magnitude
range of $\pm 0.75$ around the corresponding dSph.  The solid circles
are for the \model{NFW} (red) and \model{core13} (green) best fit
models computed for $r_{95}$ while the triangles are for $r_{50}$.

When the ratio $W_{\rm dSph}/W_{\rm sub}<1$ (all symbols {\it below}
the black horizontal line) objects need energy to transform their
halos. In these cases the binding energy of the resulting object is
larger (less negative) than that of the original untransformed
subhalo. When $W_{\rm dSph}/W_{\rm sub}>1$ (all symbols {\it above}
the black horizontal line) the binding energy of the resulting object
is smaller and such objects would have to lose energy to be
transformed into a profile that matches the observations. Regardless
of the mechanism, this transformation is energetically possible and we
do not discuss it further.

As in \citet{Penarrubia2012}, we estimate the total available energy
from SNe as
\begin{equation}
E_{\rm SN,total} = \frac{M_\star}{\langle m_\star \rangle} \xi(m_\star > 8 \Msol) E_{\rm SN},
\end{equation}
where $ \xi(m_\star > 8 \Msol) = 0.0037$ and $\langle m_\star \rangle
= 0.4 \Msol$. If the energy needed to transform a given halo is larger
than what can be delivered from SNe (i.e. the SN efficiency is larger
than 100\%), we plot it in Fig.~\ref{fig:energies} with an open
symbol. Note that for Fornax only a few open symbols exist. This means
that almost all of the candidate subhalos could be reshaped into
Fornax given its stellar content and the energy available from SNe.

For each dSph, we plot in Fig.~\ref{fig:SN_eff} the ratio of the
energy required to transform the system to the SN energy output as
estimated above, i.e. $\varepsilon_{SN}=\delta{}E / E_{\rm SN,total}$
is the SN efficiency needed to do the transformation. We thus only
plot the objects which lie below the horizontal lines in
Fig.~\ref{fig:energies}.  This analysis reveals that for each dSph,
subhalos with the right luminosities exist which can be reshaped to
follow a dark matter profile consistent with the observations.  The
required efficiencies range from $\sim 0.1\%$ up to slightly more than
100\%, depending somewhat on the extent of the region of the subhalo
that would be transformed ($r_{50}$ or $r_{95}$). Furthermore there
is always at least one subhalo that requires a SN feedback efficiency
$ \varepsilon_{\rm SN}< 0.001$.

\section{Discussion and Conclusions}
\label{sec:conclusions}

We have shown that in the two dimensional parameter space defined by
the mass and the logarithmic slope of the dark matter density profile
at $r_{-3}$, the Milky Way's dSph do not match the predictions from
dark matter only simulations. While in \citet{VeraCiro2013} we argued
that by rescaling the Milky Way mass to a (low) value of $\sim 8
\times 10^{11} \Msol$ we could circumvent the so called ``Too Big Too
Fail Problem'', we now see that halos with the right mass, do not have
the right logarithmic slope at $r_{-3}$.  However, baryons can have a
considerable effect on the shape of the dark matter density profile
through feedback \citep[e.g.][]{Governato2012}. We have shown here
that it is energetically possible at redshift of $z=0$ to reshape the
simulated halos and make them compatible with the observations.  Note
that this process would require even less energy at higher redshift,
as discussed by \citet{Amorisco2014,Madau2014}.

Hydrodynamical (cosmological) simulations of (isolated) dwarf galaxies
are now available in the literature \citep[e.g.][]{Teyssier2013,
  diCintio2014b, diCintio2014, Sawala2015}, and are more
representative of galaxy formation scenarios in $\Lambda$CDM, since
they self consistently include baryonic physics. \citet{diCintio2014},
for instance, predicts the slope in a given radial range to be
predominantly dependent of the ratio $M_\star/M_{\rm halo}$. In their
simulations an object like Fnx \citep[whose $M_\star = 4 \times 10^7
\Msol$,][]{deBoer2012}, is embedded in dark matter halo of virial mass
$M_{\rm halo} \sim 10^{10} - 10^{10.5} \Msol$. Such objects are
reported to have dark halo profile's logarithmic slopes $\alpha \in
[-1, -0.5]$ for the radial range $0.8 - 1.8$~kpc. This is inconsistent
at a 2$\sigma$ level with our measurement of the slope for Fnx at
$r_{-3}\sim 1$ kpc as can be seen from Fig.~\ref{fig:m3slopes}. Such
simulations therefore, predict slopes that are too shallow compared to
the observationally determined values \citep[see
also][]{Madau2014,Chan2015}. This suggests that the feedback scheme
used might be too strong for objects of this mass scale.

It is more difficult to make a detailed and conclusive comparison the
other dSph because the simulations' suites available in the literature
do not contain such faint objects or the ranges in which the slopes
are measured are typically larger than the observed $r_{-3}$.

The steep slopes (corresponding to high concentration) of low mass dark halos could be another manifestation of the ``Too Big to Fail'' problem \citep[however see][]{papastergis2015}, although the two issues are not necessarily strictly related, since other solutions to this problem exist \citep{Wang2012,VeraCiro2013,Brooks2014,Arraki2014}. For example, \citet{Brooks2014} argue that efficient feedback solves it, but the preliminary comparisons made above show that a careful treatment is required. We hope that our
measurements of the logarithmic slopes and masses of dSph can be used
to better constrain poorly understood processes such as the interplay
between adiabatic contraction and supernova feedback in these small
mass systems.
\\

AH and MB are grateful to NOVA for financial support. AH acknowledges
financial support the European Research Council under ERC-Starting
Grant GALACTICA-240271.
The Aquarius simulations have been run by the VIRGO consortium,
and we are very grateful to this collaboration, and particularly indebted to Volker Springel.
We are especially thankful to Else Starkenburg for the outputs of the semi-analytic code used in this Letter,
and also to Gabriella De Lucia and Yang-Shyang Li for the contributions to its development.

\bibliographystyle{apj}

\end{document}